\documentclass[twocolumn,prl,showpacs,amsmath,amssymb]{revtex4}
\usepackage{graphicx,color}

\begin{document}
\title{Quantum fluctuations of space-time}

\author{Michael~Maziashvili}
\email{maziashvili@hepi.edu.ge}\affiliation{Department of Physics,
Tbilisi
State University, 3 Chavchavadze Ave., Tbilisi 0128, Georgia \\
Institute of High Energy Physics and Informatization, 9 University
Str., Tbilisi 0186, Georgia }

\begin{abstract}
Using a \emph{gedanken} experiment providing presumably a minimal inaccuracy,
the uncertainty contributions to the space-time measurement are precisely
evaluated for clock and mirror respectively. The resulting expression of
minimal uncertainty for the
space(time) interval indicates the presence of minimal Planck scale observable
length(time). The synthesis of quantum mechanics and general relativity
predicts the UV and IR scales for Lorentz invariance violation. The influence
of background radiation on the space-time measurement is estimated.   
Based on the minimal length uncertainty relation which takes into account the
wavelength of a quantum used for distance measurement we evaluate the
cumulative factor responsible for the magnification of the space-time
fluctuation induced phase incoherence of a light propagating over a large
distance. We notice that in view of the interferometric observations the
quantum fluctuations of space-time in the braneworld model are enormously
increased if the fundamental scale is taken much below the Planck one. Present
approach to the uncertainty in distance measurement leads to new insight about the bounds of computation. The impact of the space-time fluctuations on the black hole physics is briefly emphasized.

\end{abstract}

\pacs{03.65.Ta,~ 03.67.Lx,~04.50.+h,~04.60.-m }

%03.65.Ta Foundations of quantum mechanics; measurement theory
%04.50.+h Gravity in more than four dimensions,
%Kaluza-Klein theory, unified field theories; alternative theories of gravity
%04.60.-m Quantum gravity
%03.67.Lx Quantum computation

\maketitle

The possible detection of the space-time quantum fluctuations through the cumulative uncertainty in phase and wave vector direction of the light propagating over a large distance was elaborated by several authors in the last few years \cite{LH,RTG,NDC,CND}. The smallness of the Planck length, $l_p\sim 10^{-33}$cm, characterizing the space-time foam physics takes the effect of space-time quantum fluctuations out of the range of our ordinary experiences, but the space time-foam effects can cumulatively lead to a complete loss of phase information if the emitted radiation propagated a sufficiently large distance. Since the phase coherence of light from a distant point source is a necessary condition for the presence of diffraction patterns when the source is viewed through a telescope, such observations offer sensitive and uncontroversial test for the scenarios describing the quantum fluctuations of space-time. A question of paramount importance is to define the correct cumulative factor. In papers \cite{NDC,CND} it is argued that the magnification effect of the Planck scale fluctuations during the light traveling over a long distance is overestimated in \cite{LH,RTG} because of improper definition of the cumulative factor. First let us try to clarify this point.                

That the space-time undergoes quantum fluctuations can be simply demonstrated
by analyzing a \emph{gedanken} experiment for space-time measurement \cite{SW}. Due
to importance of a numerical factor characterizing amplitude of the space-time
fluctuations we shall try to be as precise as possible in analyzing of
\emph{gedanken} experiment. Our measuring device is composed of a spherical
clock localized in the region $\delta x=2r_c$ and having a mass $m_c$, which
also serves as a light emitter and receiver, and a spherical mirror of radius
$l/2$ and mass $m$ surrounding the clock, ($r_c$ denotes the radius of the
clock), Fig.1. We are measuring a distance $l$ by sending a light signal to the
mirror under assumption that at the moment of light emission the centers of
clock and mirror coincide. However, quantum uncertainties in the positions of
the clock and mirror introduce an inaccuracy $\delta l$ during the
measurement. (Throughout this paper we set $\hbar=c=k_B=1$).
\begin{figure}[t]

\includegraphics{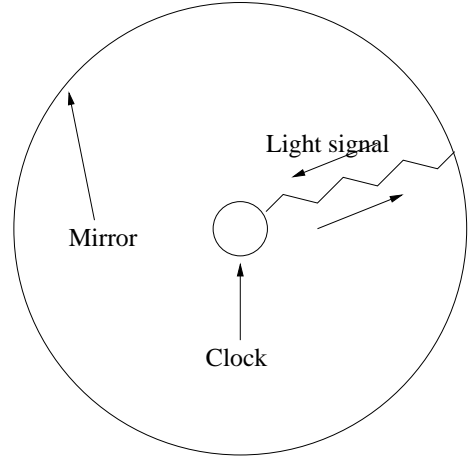}\\
\caption{ The device for distance measurement.}
\end{figure}
Assuming the minimal uncertainty the spread in velocity of the clock
can be found as $ \delta v_c = \delta p_c/m_c =1/2m_c\delta x$ and analogously for the mirror as $ \delta v_m  =1/2(m+m_c)l$. After the
time $t = l$ elapsed by light to travel along the closed path
clock-mirror-clock the total uncertainty during the measurement
takes the form \[
\delta l = \delta x +{ t\over 2m_c\delta x} +{ t\over 2(m+m_c)l}~, \] which after
minimization with respect to $\delta x$ gives
\begin{equation}\label{minim}\delta x=\sqrt{{l \over 2 m_c}}~,~~~\Rightarrow~~~\delta l_{min}=\sqrt{{2l
\over  m_c}}+{1\over 2(m+m_c)}~.\end{equation} This uncertainty diminishes with increasing masses of the mirror and clock. But the masses are limited by the requirement the device not to collapse into the black hole. In other words the size of the clock(mirror) should be greater than twice its gravitational radius. Following this discussion and using the expression of gravitational radius \[r_g=(Al_p^{2+n}m)^{{1\over 1+n}}~,~~~~~~ A={8\Gamma\left({3+n\over 2}\right)\over (2+n)\pi^{{n+1\over
2}}}~, \] where $n$ denotes the number of extra dimensions, one simply gets
\begin{equation}\label{uncleng}\delta l_{min}=2^{{3+2n\over 3+n}}A^{{1\over
3+n}}\left (l_p^{2+n}l \right)^{{1\over 3+n}}+2^nAl_p^{2+n}l^{-(1+n)}~.
\end{equation} From this equation it follows immediately that there exists the
minimal observable length of the order of $\sim l_p$. From Eq.(\ref{uncleng}) one readily gets the minimal uncertainty
in time measurement, $\delta t_{min}$, by replacing $l_p$ and $l$ with $t_p$
and $t$ respectively. The latter term in
Eq.(\ref{uncleng}) becomes negligible in comparison with the first one if
$l\gg l_p$. In what follows we assume the condition $l\gg l_p$ that results in the following expression for the minimal length uncertainty    
\begin{equation}\label{minlengunc}\delta l_{min}\approx 2^{{3+2n\over
3+n}}A^{{1\over 3+n}}\left (l_p^{2+n}l \right)^{{1\over 3+n}}~.\end{equation}
Loosely speaking, in Eq.(\ref{uncleng}) one can consider the first and second
terms as the uncertainties contributed to the measurement by the clock and
mirror respectively. Due to maximal symmetry of the measuring device
considered here one can hope it provides minimal uncertainty in length
measurement.

The above discussion to be self-consistent let us notice that due to
Eq.(\ref{minlengunc}) the wavelength of photon used for a measurement may be
known with the accuracy $\delta\lambda\sim (l_p^{2+n}\lambda)^{1/(3+n)}$ that
results in additional uncertainty \begin{equation}\label{totalucerlam}\delta
l_{min}\sim\left (l_p^{2+n}l\right)^{{1\over
(3+n)}}+{l\over\lambda}\delta\lambda \sim\left (l_p^{2+n}l\right)^{{1\over
(3+n)}}+{l_p^{{2+n\over 3+n}}l\over \lambda^{{2+n\over 3+n}}}~.
\end{equation}(For the sake of simplicity we assume that the wave length of
the photon is not affected by the gravitational field of the clock). For $\lambda$ can not be greater than $l$ in order to measure this distance the latter term in Eq.(\ref{totalucerlam}) can be minimized by taking $\lambda\sim l$. In this case the latter term becomes comparable to the first one and therefore justifies the Eq.(\ref{minlengunc}). In the case when $\lambda^{{2+n\over 3+n}}\ll l^{{2+n\over 3+n}}$ the second term in Eq.(\ref{totalucerlam}) dominates and correspondingly the minimal uncertainty and  metric fluctuations take the form \begin{equation}\delta l_{min}\sim {l_p^{{2+n\over 3+n}}l\over \lambda^{{2+n\over 3+n}}}~,~~~~~~\delta g_{\mu\nu}\sim \left({\l_p\over\lambda}\right)^{{2+n\over 3+n}}~. \label{fluc} \end{equation}

In general the minimal length uncertainty has the form
\begin{equation}\label{alpmin}\delta l_{min}\sim
l_p^{\alpha}l^{1-\alpha}~,\end{equation} where the parameter $\alpha$
specifies different scenarios: $\alpha=2/3$ \cite{SW},  $\alpha=1/2$
\cite{AC}, $\alpha=1$ \cite{MTW} and, as it shown in \cite{Ma}, for ADD
braneworld model \cite{ADD} with the size of extra dimensions much exceeding
the fundamental Planck scale, $\sim (\mbox{TeV})^{-1}$,~ $\alpha=(2+n)/(3+n)$
\cite{note1}. From Eq.(\ref{alpmin}) one simply gets that the minimal length
uncertainty and metric perturbations when one uses for measurement the quantum
with wavelength $\lambda$ take the form  \begin{equation}\delta
l_{min}\sim\left( {l_p\over \lambda}\right)^{\alpha}l~,~~~~~~\delta
g_{\mu\nu}\sim \left({\l_p\over\lambda}\right)^{\alpha}~. \label{fluc}
\end{equation} So that the space-time perturbation during the measurement when
one uses the quantum with a wavelength $\lambda$ to probe the space-time
region with a linear size $l$ is the greater the shorter the wavelength of the
quantum used for measurement is. Therefore, for the amplification factor one
gets \begin{equation}\label{ampliffact}{\delta l\over \delta\lambda}\sim
{l\over\lambda}~,\end{equation} which is in complete agreement to what is used
in \cite{LH,RTG}. So the key point missed in \cite{NDC,CND} is that the
dependence of space-time perturbations on the wavelength of quantum used for
the measurement is not taken into account.

As it is discussed in \cite{LH,RTG} due to amplification factor given by Eq.(\ref{ampliffact}) the distant compact radiation sources should not produce the normal interference patterns that are often observed. Hence, if the experimental results analyzed in \cite{LH,RTG} can be taken to be trustworthy then we must face the challenge of finding a new, self-consistent, formulation of the fundamental laws that agrees with experiment for the derivation of space-time quantum fluctuations is based on accepted principles of quantum mechanics and general relativity. Let us stress that the brane induced quantum gravitational fluctuations evaluated in \cite{Ma} are unacceptably enhanced by taking the fundamental scale much below the Planck one. In view of the papers \cite{LH,RTG} one simply finds that in the case of braneworld model with $\sim$TeV fundamental scale the theory fails at least by $10$ orders of magnitude. Even if one takes the amplification factor $(l/\lambda)^{1-\alpha}$ \cite{NDC,CND}, which seems to be not correct as it is discussed above, one finds that the braneworld model with $\sim$TeV fundamental scale fails by order of magnitude if only phase decoherence of the light detected by the telescope is taken into account and by $7$ orders of magnitude if the fluctuations in direction of the wave vector is also taken into account.   

In detecting the space-time foam one has to take into account the side factors
as well that can overlap this effect. One of such factors can be the
space-time perturbations caused by the background radiation. Since we are
using a bit modified measuring device let us first revise the corresponding
result obtained in \cite{Ma}. We restrict ourselves to the case $n=0$. Let us insert our measuring device into the
background radiation with a temperature $T$. Because of the
background temperature the device acquires a mean velocity
$\sim\sqrt{T/(m+m_c)}$. Assuming the gravitational radius
of the device is not changed significantly due to background
radiation and repeating the above discussion the expression of minimal
uncertainty takes the form 

\[\delta l_{min}\sim l_p^{2/3}l^{1/3}+l_p\sqrt{lT}~.\] One sees that the
present measuring device reduces significantly the influence of CMB on the
length uncertainty relative to that one considered in \cite{Ma}. Namely, for
$T=2.7$K the latter term in this equation becomes appreciable for distances
$l^{1/6} \gtrsim   10^{11}$cm$^{1/6}$ much exceeding the present size of the
observable universe. In general, when the photon with the wavelength $\lambda$
is used for the measurement one gets  \[\delta l_{min}\sim {l_p^{2/3}\over
\lambda^{2/3}}l+l_pl\sqrt{{T\over \lambda}}~.\] Correspondingly the effect of
the CMB becomes negligible if $\lambda^{1/6}\ll 1/l_p^{2/3}T^{1/2}$. In
experiments considered in \cite{LH, RTG,NDC,CND} $\lambda\sim \mu$m and therefore this condition is satisfied
with a great accuracy.

Now let us turn to the issue of computation for which the Eq.(\ref{fluc}) can
provide new insight. As it was shown in \cite{Ba} the results of Salecker and
Wigner \cite{SW} provide quite strong constraints on the ultimate capability
of the computer. This idea was further developed in \cite{Ng}. Let us consider
a simple computational system as a cube $l^3$ filled with the bits $l_b^3$
inside of which photon is bouncing representing therefore a bit-operation
process. Using the Heisenberg uncertainty relation one simply concludes that the
 lifetime of the computer, i.e., the time during which it will be confined to
 the region $l^3$ is given by \[T\sim l^2M~,\] where $M$ is the mass of the
 computer. The computer not to collapse $l$ should be greater than its
 gravitational radius $\sim l_p^2M$ resulting therefore in the notion of
 maximal lifetime and mass of computer with a given size $l^3$ \[ M_{max}\sim{l\over l^2_p}~,~~~~~~T_{max}\sim {l^3\over
 l_p^2}~.\] In view
of the existence of minimal insurmountable uncertainty in distance measurement
the size of the bit cannot be smaller than $\delta l_{min}$. Simply speaking we need to write an information in the bit, read
and move it to another bit. So that inside the region $l^3$ we need to operate
with the quanta having the wavelength comparable to the size of a bit.
Correspondingly from Eq.(\ref{fluc}) one can evaluate the size of bit
\begin{equation}\lambda\sim\left
({l_p\over\lambda}\right)^{\alpha}l~,~~~~\Rightarrow
~~~~\l_b \sim(l_p^{\alpha}l)^{1/( 1+\alpha)}~,\end{equation} the minimal
time for bit-operation and the maximal number of bits
\begin{equation}t\sim(l_p^{\alpha}l)^{1/( 1+\alpha)}~,~~~~  N\sim\left
({l\over l_p}\right)^{3\alpha/(1+\alpha)}~.\end{equation} In this case for
maximal number of operations one gets \[N{T_{max}\over t}\sim  \left ({l\over
l_p}\right)^{(2+6\alpha)/ (1+\alpha)} ~.\] So this is the maximal number of
operations the computer can accomplish in a given region with linear size $l$.
In the above considered partition of $l^3$ into the bits $l_b^3$ the energy
per bit is \[E_b={M_{max}\over N}\sim {1\over l_p^{{2-\alpha\over
1+\alpha}}l^{{2\alpha-1\over 1+\alpha}}}~.\] Correspondingly from time-energy
uncertainty relation one can say that the minimum time for bit-operation
should be $E_b^{-1}$ which is less than $l_b$ used above and leads to the
Margolus-Levitin \cite{MaLe} limit for maximum number of operations the
computer can perform in region $l^3$ given by $\sim T_{max}M_{max} $. So in
general there are two time scales for the bit, $l_b$ and $E_b^{-1}$. In order to arrive at the Margolus-Levitin limit unambiguously these two
scales should coincide. Assuming this condition, $l_b= E_b^{-1}$,one gets
\[{M\over E_b}=N~,~~l_b={l\over N^{1/3}}~,~~\Rightarrow~~ E_b={M^{1/4}\over
l^{3/4}}~, \] which by taking into account $l_b\sim
(E_bl_p)^{\alpha}l$ implies \[M\lesssim {l^{{3\alpha -1 \over 1+\alpha}}\over
l_p^{{4\alpha \over 1+\alpha}}}~. \] So, for $\alpha < 1$ one sees that this
upper bound on the mass is less than $M_{max}\sim l/l_p^2$ and therefore the
Margolus-Levitin limit is not attainable in this case for computer having the
mass $M_{max}$. The case $\alpha=1$ is exception allowing the Margolus-Levitin
limit for maximum number of operations no matter what the mass of computer is. 

It is of interest to analyze the minimal length uncertainty expression from
the standpoint of Doubly-special relativity suggesting the Planck length as a
second relativistic invariant besides the speed of light. This argument can be
motivated by the Generalized Uncertainty Principle (GUP) ensuring the presence of
minimal particle localization region with linear size of the order of Planck
length \cite{AC1}, which in turn can also be derived by combining the
uncertainty relation with gravitation \cite{AS}. (The GUP derived via the quantum corrected gravitational potential \cite{Ma1}
as well as for the ADD braneworld model \cite{MMN} exhibits the linear size
for the minimal localization region of the quantum to be of the order of
Planck length as well). From the Eq.(\ref{alpmin}) one sees that if $l_p$ is invariant
of the theory and $l$ transforms with respect to the special relativity then
the Lorentz invariance is violated at the scale $\delta l_{min}$ and vice
versa. By taking $l\sim $TeV$^{-1}\sim 10^{-16}$cm, which can be considered as
a minimal length scale for which the Lorentz invariance is an experimental
fact, one finds $\delta l_{min}\sim 10^{-16(1+\alpha)}$. So one can say that
the transition from special relativity to the doubly-special regime takes
place beneath the scale $  \lesssim 10^{-16(1+\alpha)}$cm. When $\delta
l_{min}$ approaches $\sim 10^{-16}$cm, the length scale $l\sim 10^{(33\alpha
-16)/( 1-\alpha)}$cm determines the upper bound for the special relativity.
from this expression one sees that for $\alpha=1/2$ \cite{AC} the upper bound
for Lorentz invariance becomes unacceptable small. For $\alpha=2/3$ \cite{SW}
the upper bound for Lorentz invariance is about $\sim$pc. This result is
interesting in that it indicates every Lorentz invariant theory including the
Maxwell electrodynamics should contain this characteristic scale beyond which
the modification of the theory takes place. Correspondingly, the consistent
treatment of the phase incoherence accumulation for electromagnetic wave
traveling the distance much exceeding the length scale $\sim$pc requires the
modified theory to be known and used respectively.

Finally, following the paper \cite{Ma}, let us briefly emphasize the impact of space-time fluctuations on the black hole physics. The presence of unavoidable uncertainty in length measurement provided by all the space-time foamy models listed above results in fluctuations of the black hole thermodynamics simply because the gravitational radius can not be known exactly. Because of these fluctuations one can no longer argue that the information about the initial state of the body that collapsed to form the black hole is lost during the black hole evaporation. But it is principally impossible to keep track of these fluctuations that remains open the question about the unitarity in evolution of the initial state during the black hole formation and subsequent evaporation. The quantum corrections to the black hole entropy obtained in various scenarios are indiscernible because of these
fluctuations. Fluctuations near the Planck scale become of the order of thermodynamic quantities themselves and therefore destroy the thermodynamical picture of the black hole. So that one should be cautious about the drawing sweeping conclusions based on the black hole thermodynamics near the Planck scale. For more details see \cite{Ma}.   

To summarize, on the bases of \emph{gedanken} experiment depicted in Fig.1 we
have precisely evaluated the minimal uncertainty in space-time measurement.
The expression of minimal uncertainty obtained in this way immediately results
in minimal observable length(time) of the order of Planck length(time). We
have specified the dependence of minimal length uncertainty on
the wavelength of the quantum used for the measurement. On the bases of this
relation the cumulative factor describing the phase and wave-vector
uncertainties for the wave traveling over a large distance as well as bounds
on the computation are evaluated. On the one hand the synthesis of uncertainty
relation with gravitation leads to the GUP \cite{AS} which
shows that there is a minimal particle localization limit of the order of
Planck length and therefore motivates the Planck length as a second invariant
of the theory together with the speed of light \cite{AC1}. On the other hand combining
together the uncertainty relations and general relativity one gets the minimal unavoidable
uncertainty in space-time measurements which by taking into account the Planck
length as an invariant exhibits the Lorentz invariance region from $\sim
10^{-26}$cm to $\sim 10^{18}$cm.  So one sees that one of the most striking results coming from synthesis of quantum mechanics
and general relativity may be the prediction of Lorentz violation scales.
Correspondingly, one can not make some sweeping conclusions on the bases of
papers \cite{LH, RTG} about the discrepancy between theory and experiment as further
progress has likely to be achieved to provide new insights into the
fundamental theories.                     
          
\vspace{0.2cm}
\begin{acknowledgments}
The author is greatly indebted to David Langlois for cordial hospitality at APC (Astroparticule et Cosmologie, CNRS, Universit\'e Paris 7) and IAP (Institut d'Astrophysique de Paris), where this work was done. The work was supported by
the \emph{INTAS Fellowship for Young Scientists}, the
\emph{Georgian President Fellowship for Young Scientists} and the
grant \emph{FEL. REG. $980767$}.
\end{acknowledgments}

%%%%%%%%%%%%%%%%%%%%%%%%%%%%%%%%%%%%%%%%%%%%%%%%%%%%%%%%%%%%%%%%%%%%%%%%


\begin{thebibliography}{9}
\bibitem{note1}
It is easy to see that the measurement device proposed here gives the same
result for the ADD braneworld model under assumption $l\gg l_p$. Namely, if
$l$, the size of the device, is less than the size of extra dimensions, $L$,
one gets for minimal length uncertainty the Eq.(\ref{uncleng}). If $l$ is greater
than $L$ but the size of clock is still less than $L$ the second term in
Eq.(\ref{uncleng}) apperas with $n=0$. If the size of clock is greater than $L$ one
gets the Eq.(\ref{uncleng}) with $n=0$, i.e., the pure four-dimensional
result. By the way one has to keep in mind the difference between the four and
higher dimensional Planck scales.
\end{thebibliography}
\end{document}